# Electrochemical 3D Printing of Ni-Mn and Ni-Co alloy with FluidFM

Chunjian Shen[1,2], Zengwei Zhu[1*], Di Zhu[1], Cathelijn van Nisselroy[2], Tomaso Zambelli[2], Dmitry Momotenko[3]

[1] College of Mechanical and Electrical Engineering, Nanjing University of Aeronautics and Astronautics, P.R. China
[2] Laboratory of Biosensors and Bioelectronics, ETH Zürich, Gloriastrasse 35, 8092, Zurich, Switzerland
[3] Department of Chemistry, Carl von Ossietzky University of Oldenburg, Oldenburg, D-26129, Germany

Corresponding author: zhuzw@nuaa.edu.cn,

**Abstract**

Additive manufacturing can realize almost any designed geometry, enabling the fabrication of innovative products for advanced applications. Local electrochemical plating is a powerful approach for additive manufacturing of metal microstructures; however, previously reported data have been mostly obtained with copper, and only a few cases have been reported with other elements. In this study, we assessed the ability of fluidic force microscopy (FluidFM) to produce Ni-Mn and Ni-Co alloy structures. Once the optimal deposition potential window was determined, pillars with relatively smooth surfaces were obtained. The printing process was characterized by printing rates in the range of 50-60 nm/s. Cross-sections exposed by focused ion beam showed highly dense microstructures, while the corresponding face scan with energy-dispersive X-ray spectroscopy (EDX) spectra revealed a uniform distribution of alloy components.

## 1. Introduction

Compared to pure metals, alloys often exhibit superior performance, offering higher strength, better ductility, and increased corrosion resistance, making them a material of choice for many applications. Recently, microscale processing of metallic materials has attracted significant attention. Microdevices made of alloys are key components in industrially manufactured microsystems [1-5], such as mechanical timepieces, μ-gears, interconnects, and microprobes. Most of these microstructures are fabricated using standard or UV versions of LIGA (lithographie, galvanoformung, abformung) techniques. From the perspective of advanced industrial applications, alloys are expected to play a major role in complex 3D microstructures such as photonic crystals [6, 7], interconnects [8, 9], and biosensors [10, 11]. However, common techniques such as LIGA or UV-LIGA are not always well suited for preparing such complex 3D designs.

Additive manufacturing (AM) can realize almost any designed geometry, enabling the fabrication of innovative products for advanced applications. At the micrometer scale, some representative additive manufacturing techniques have shown potential for producing complex metal structures via direct ink writing (DIW) [12, 13], electrohydrodynamic printing ((EHD) [14, 15]), and focused electron/ion beam-induced deposition (FEBID/FIBID) [16, 17]. Alloys are generally subdivided into solid solutions (atoms of other components integrated together in a single lattice), metallic compounds (components ordered in a particular lattice), and simple mixtures. DIW and EHD rely on metal nanoparticles, which implies that they can target alloys with mechanically-mixed metal nanoparticles. Nevertheless, data reported in the literature regarding these two techniques only involve microstructures built with a single metal, where the prepared metal has a low strength and is non-conductive because nanoparticles can be surrounded by a less conductive organic impurity. In contrast, FEBID/FIBID is based on beam-induced dissociation of suitable precursor molecules. Successive reports on preparing alloys with these types of techniques have shown potential for fabricating a solid solution alloy of Co-Pt [18] and a metallic compound alloy of Pt-Si [19]. However, the metal purity of the obtained microstructures was as low as 10–20 at. %, with a high content of other elements inside the metal, such as C, O, and Si. Despite the intrinsic strengths of focused ion/electron beam methods, particularly in terms of resolution and fabrication complexity, alloy microstructures with low contaminant levels remain a challenge.

In this regard, alloys prepared by electrodeposition have the advantage of better overall quality, particularly in terms of purity and composition uniformity. This implies that additive manufacturing techniques based on electrochemical (EC) printing approaches have high potential for the preparation of high-quality alloy microstructures. The family of EC printing techniques includes microanode-guided electroplating (MAGE) [20], meniscus-confined electrodeposition (MCED) [21], fluidic force microscopy (FluidFM) printing [22], electrohydrodynamic redox printing (EHD-RP) [23], and a range of techniques based on scanning ion conductance microscopy (SICM) [24]. Among these techniques, MAGE, MCED, and EHD-RP have prepared alloys and multicomponent prints [23, 25, 26], which have shown that EC printing has a high potential for manufacturing alloy microstructures.

Simultaneously, local EC plating of alloys can, in some cases, be complicated either because of the requirement of processing two solutions (at least) containing each metal salt separated in two different reservoirs (approach of EHD-RP) or the two metal salts in the same reservoir (approach of MAGE and MCED). In the first case, the challenge is the engineering of a system with two synchronized dispensers, which in some cases may be less optimal for achieving uniform component distribution [23]. In the second case, the challenge is to control the simultaneous EC plating of the two metal ions; however, printing of alloy microstructures using MAGE and MCED can also be achieved [25, 26]. However, these two approaches are often viewed as being limited to the fabrication of 3D structures with high complexity. In this context, FluidFM technology has been shown to be the most advanced in terms of printing complex designs [27, 28], which has printed an impressive 1:70,000 replica of Michelangelo's David, but so far has not been extended to printing alloys or micromaterial structures. Given that FluidFM cantilevers are equipped with only one embedded microchannel, this task can be tackled by electroplating from a mixture of metal ions.

In this study, we extend the capacity of metal printing by using FluidFM for simultaneous Ni-Mn and Ni-Co electrodeposition. The strategy for microfabrication was adapted to 3D printing with FluidFM using a well-established electroplating processe [27, 28]. The printed materials were characterized using scanning electron microscopy (SEM) and energy-dispersive X-ray spectroscopy (EDX), and it was found that the structures had a relatively smooth surface and a uniform distribution of alloy components.

## 2. Material and Methods

A three-electrode cell was used for EC printing with the FluidFM. An Si substrate (1.3 × 1.3 cm2) coated by a 3 nm Ti adhesion layer and sputtered by a 25 nm Au top layer, was used as the working electrode, and a Pt electrode and an Ag/AgCl electrode were used as the counter electrode and quasi-reference counter electrode, respectively. The solutions for alloy electrodeposition were prepared according to previous references [29-31]. In particular, the Ni-Mn electrodeposition solution consisted of 1 M NiSO4, 0.11 M MnSO4, and 0.49 M H3BO3. The Ni-Co electrodeposition solution consisted of 1 M NiSO4, 0.09 M CoSO4, and 0.49 M H3BO3. The supporting electrolyte consisted of 0.57 M H3BO3 and 0.04 M Na3C6H5O7 at a pH of 6. All chemical reagents were purchased from Sigma-Aldrich.

A Palmsense3 potentiostat (Palmsens B.V., The Netherlands) was used to perform electrochemical voltammetric tests for macroscopic tracking of the electrodeposition process from the two solutions. The scan rate and range of the cyclic voltammogram were 0.02 V/s and 0 V to −1.5 V, respectively.

A FluidFM BOT (Cytosurge AG, Switzerland) equipped with a microchanneled cantilever probe with a 300 nm aperture at the pyramidal apex was used to print the alloy pillars. This cantilever was used as a local source of metal ions in an electrochemical cell, leading to localized electrodeposition directly under the tip aperture [22]. A feedback mechanism was employed to detect the growth of the deposit, thereby realizing voxel-by-voxel 3D printing [27, 28]. The voxel height for the pillar printing was 250 nm. After printing, all the samples were characterized using scanning electron microscopy (SEM; Regulus 8200, Hitachi, Japan), and EDX analysis was conducted at an acceleration voltage of 30 kV. FIB-SEM (Helios G4 PFIB, FEI, America) was used to mill the cross-sections of the pillars with a current of 30 pA. A protective layer containing 95% Pt and 5% C was sputtered onto the pillar surface before milling. The EDX face scan of the cross section was performed at an acceleration voltage of 30 kV. The microstructure of the pillars was characterized using transmission electron microscopy (TEM; JEM-2100F, JEOL, Japan), and an EDX scan was conducted at an acceleration voltage of 200 kV.

## 3. Results and Discussions

The principles of microscale 3D printing techniques via electrodeposition, including MAGE, MCED, and FluidFM, are shown in Fig. 1. The common concept underlying these three techniques is based on (i) synthesizing the solid metal in-situ electrochemically, and (ii) performing this process locally. In MAGE (Fig.1a), this is achieved by focusing the electric field in the gap between the substrate and the mobile anode. In MCED (Fig.1b), the process is restricted by the size of the droplet cell itself, whereas in FluidFM (Fig.1c), the metal ions are only supplied locally. This is achieved by manipulating small liquid volumes using a microchanneled cantilever, which acts as a local dispensing pipette. This means that the electrodeposition electrolyte is locally dispensed in the supporting electrolyte first, then diffuses to the working electrode, and finally induces electrodeposition, which makes the electrodeposition process of FluidFM much different from that of MAGE and MCED.

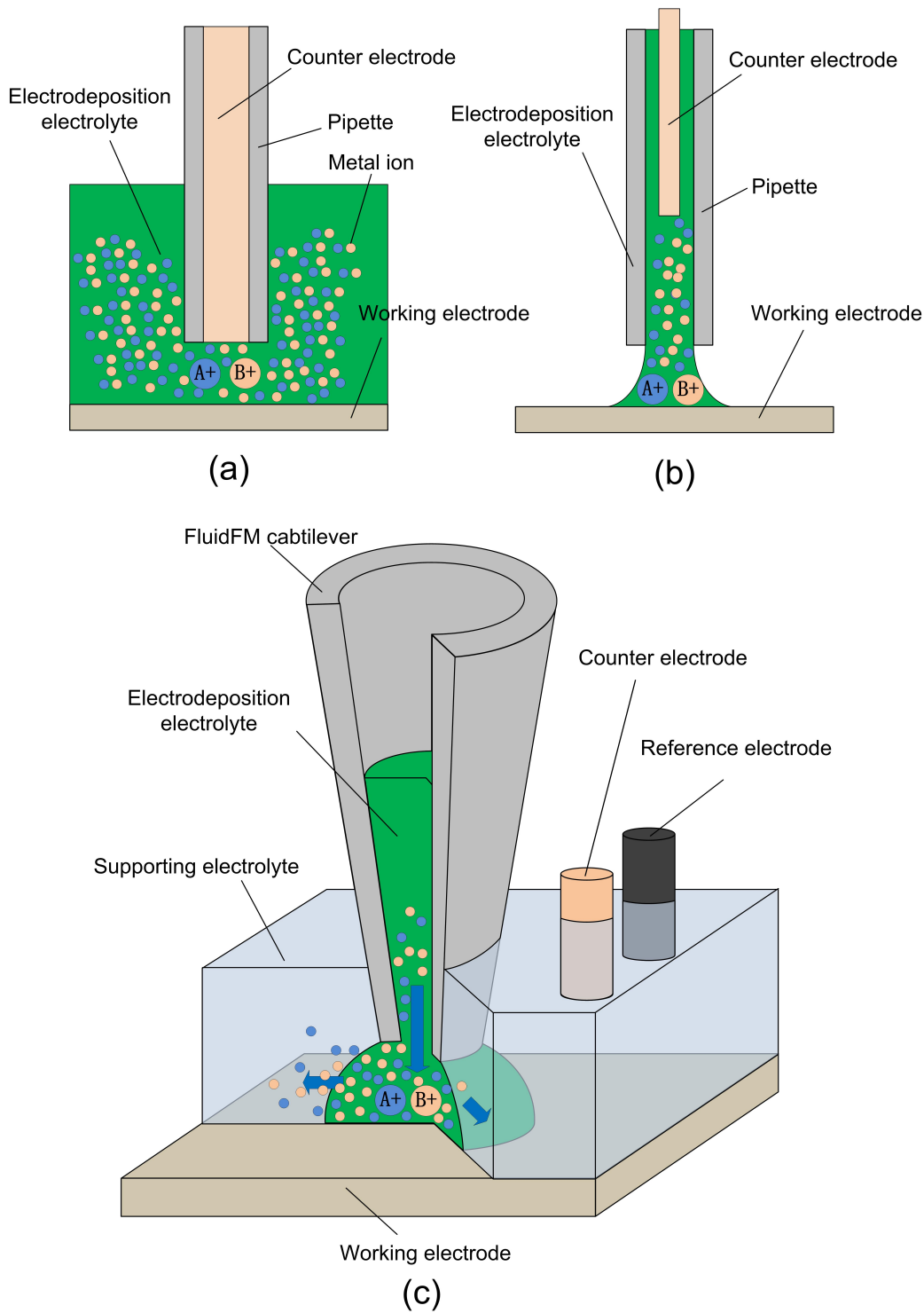

***Fig.1.*** *Schematic diagram of alloy electrodeposition via **(a)** MAGE, **(b)** MCED and **(c)** EC printing with FluidFM technology*

To study alloy electrodeposition with a local dispensing process and to determine the most suitable reductive potential range for FluidFM, macroscopic electrochemical experiments were carried out in a three-electrode cell (Fig.2). Initially, the cyclic voltammetry (CV) curves of the supporting electrolyte were recorded. As shown in Fig. 2 (a), the voltammogram is almost a flat line, indicating the absence of any significant EC reactions and confirming that the electrolyte is reasonably inert within the chosen potential window. Subsequently, 10 μL of the electrodeposition solution was injected onto the working electrode surface into 50 ml of the supporting electrolyte. This leads to a drastic change in the voltammogram, indicating strong EC reduction processes at voltages below -0.9 V. The voltammetric behaviors of both Ni-Mn and Ni-Mn solutions are similar, probably because of the main contribution to the current given by $Ni^{2+}$ ions (present in higher concentration), coupled with a hydrogen evolution reaction, which readily occurs on the deposited metal features.

Electroplating was also confirmed visually by the color change of the working electrode at different voltages. Similar to the CV curves, 10 μL of the Ni-Mn and Ni-Co electrodeposition solutions were injected into the supporting electrolyte. During these tests, constant potentials of −1.2 V, −1.3 V, and −1.4 V were applied at 3 s, respectively. As expected, the intensity of the color change due to the quantity of the deposited metal was observed with decreasing potential (Fig. 2b and 2c). When the potential was below −1.4 V, a strong hydrogen evolution reaction on the substrate was observed, resulting in intense bubble formation. This can significantly impair the use of FluidFM for printing because the bubbles can affect the laser-based atomic force microscopy (AFM)-like force feedback of FluidFM. Therefore, a plating window from −1.3 V to −1.4 V for Ni-Mn and Ni-Co was chosen for EC printing.

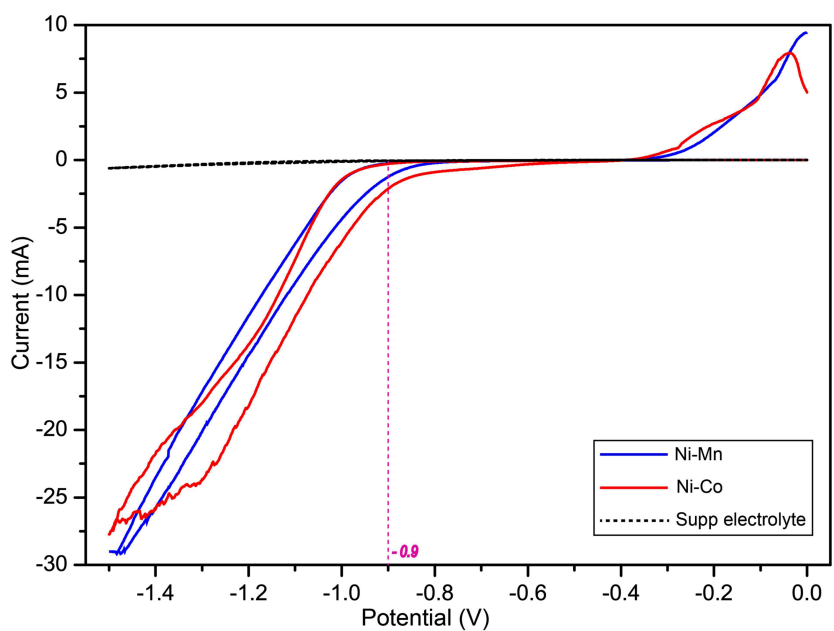

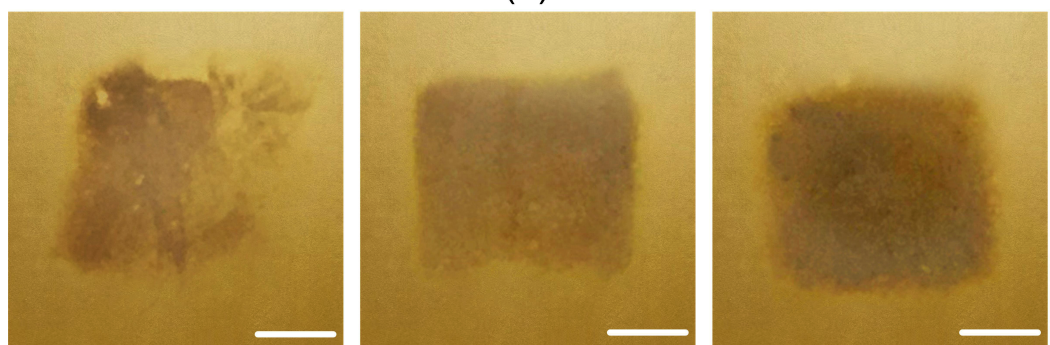

-1.2 V , 3 s     -1.3 V , 3 s     -1.4 V , 3 s

(b)

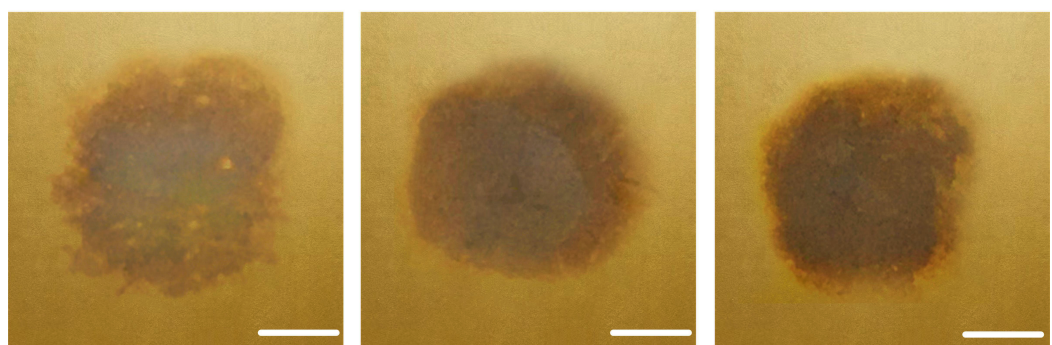

-1.2 V , 3 s     -1.3 V , 3 s     -1.4 V , 3 s

(c)

***Fig.2.*** *results of electrochemical tests: **(a)** voltammograms measured in the supporting electrolyte (black dotted line) as well as with the addition of Ni-Mn (blue) and Ni-Co (red) salt solutions. Optical images acquired after constant potential deposition tests with of Ni-Mn **(b)** and Ni-Co **(c)** solutions. The scale bar is 2 mm.*

The chosen salt solutions and the voltage window were then employed to fabricate micropillars, which are the simplest 3D structures. Micropillar printing is relatively simple but allows us to gain details about the printing process such as, potential problems with nozzle clogging, printing speed, structure surface morphology, and internal microstructure of the printed features. All the printed pillars were 10 μm high, corresponding to the maximum range of the positioning system of the FluidFM instrument. The working electrode potential and pressure inside the microchannel are two parameters that can be tuned during the AM process. The potential determines the rate of the electrochemical metal plating process, and the pressure determines the mass transport rate of metal ions ejected from the tip aperture. For printing of Ni-Mn pillars the working electrode potentials of −1.35 V, −1.37 V, and −1.39 V were applied while the pressure was kept at 8 mbar. As shown in Fig.3 (a), pillars with relatively smooth surfaces are obtained. A few abnormal bulges were locally distributed on the surface. The diameter of the pillar decreased with an increase in the applied potential, and the diameter difference between the bottom and top of the pillar was reduced. This is attributed to the collection efficiency of metal ions, which is the interplay between the electrodeposition kinetics and mass transport of metal ions in the supporting electrolyte originating at the tip aperture. Similar to a wall-jet electrode configuration, at higher cathodic potentials, electrodeposition is faster, the collection efficiency is higher, and the pillars decrease in diameter. A similar process was observed for the electrodeposition of copper [28]. The influence of collection efficiency is also observed during the printing process. The electrodeposition happens under the tip aperture. The collection efficiency cannot be 100%. The number of metal ions originating at the tip aperture increases with time. More metal ions diffused and induced deposition on the printed voxels. Therefore, the diameter of the Ni-Mn pillar increased with an increase in the voxel, as shown in S1. The printing speed increased with decreasing potential so that the average growth rates at −1.35 V, −1.37 V, and −1.39 V were approximately 52.3 nm/s, 55.8 nm/s, and 60.6 nm/s, respectively.

For printing the Ni-Co pillars, the potential was maintained at 1.35 V and pressures of 6, 8, and 10 mbar were applied. As shown in Fig.3 (b), the pillars also exhibited a smooth surface morphology. The pillar diameters increased with increasing pressure. The printing speed in this case is almost constant regardless of the pressure and reaches 50.3 nm/s, 50.8 nm/s, and 51.2 nm/s for 6 mbar, 8 mbar, and 10 mbar, respectively. The increase in diameter at higher pressures is also a consequence of the wall-jet configuration, as previously reported [28].

To determine the composition of the pillars, EDX experiments (Fig.3(c)) were performed. Besides the expected Ni, Mn, and Co peaks, other elements, such as O and C were also found in the spectra, which can be attributed to oxidation and to organic matter being present in the air during post-printing handling. For Ni and Mn, the wt% proportion of metals was 13.29, and 1.53 for Ni and Co, with Ni being the major alloy component. The EDX analysis of all pillars showed that the potential and pressure did not affect the alloy composition (Fig.3(d)). The metal content of the pillars is similar to the typical electrodeposited thin films (from these salt solutions), in which the wt% proportion for Ni and Mn is typically more than 9 [32, 33], and the wt% proportion for Ni and Co is approximately 1.5 ~ 4 [29, 34].

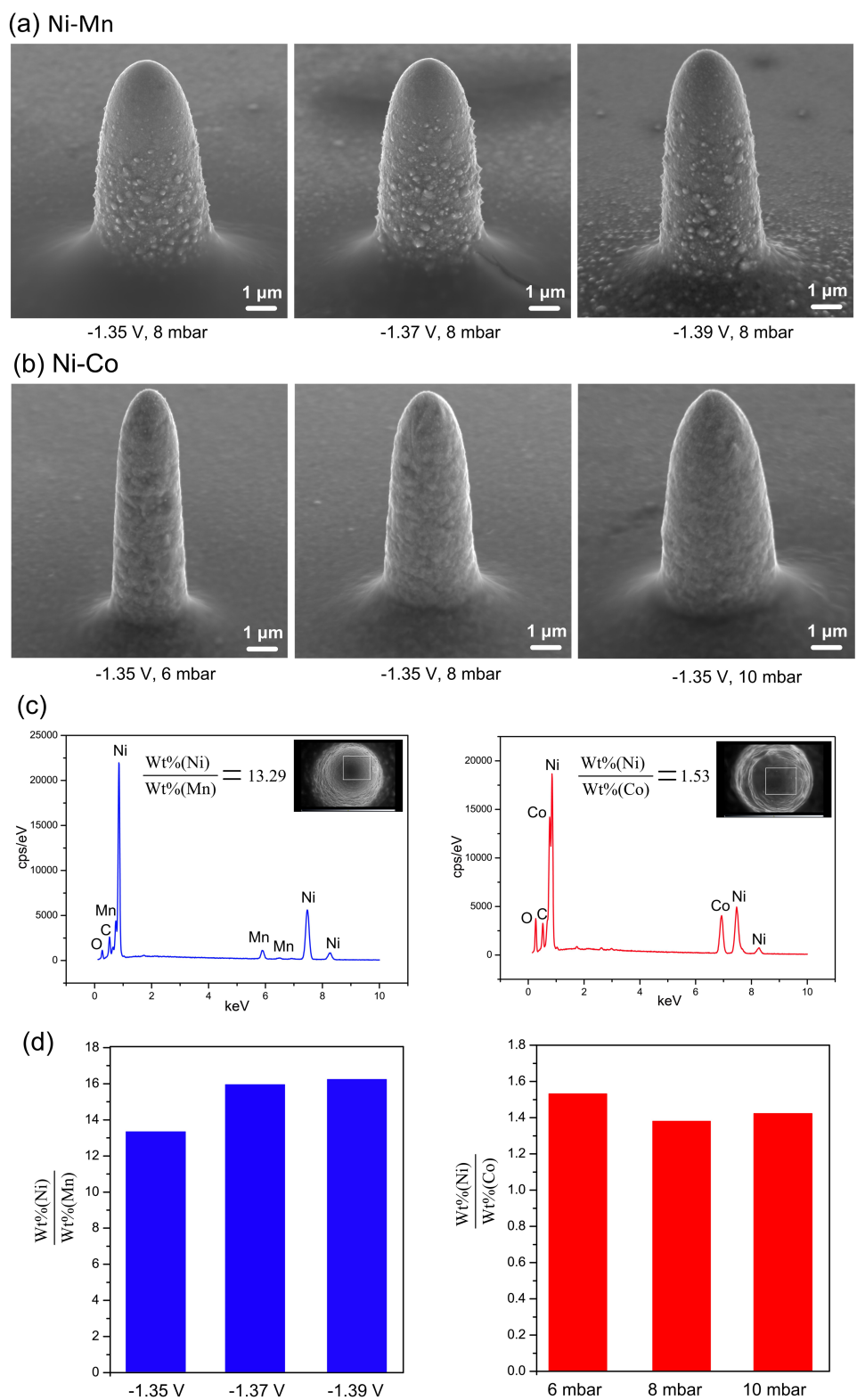

***Fig.3.*** *Ni-Mn **(a)** and Ni-Co **(b)** printed pillars. **(c)** EDX spectra of pillars, **(d)** the composition (wt%) proportion of metals in pillars at varying voltages for Ni and Mn (left) and pressures for printing Ni and Co (right).*

To confirm the dense nature of the electrodeposits, cross-sections of the Ni-Mn and Ni-Co pillars were obtained using focused ion beam (FIB) milling, as shown in Fig.4(a). FIB was employed here to expose the inner structure of the printed features while EDX face scans enabled the distribution of Ni, Mn, and Co on the cross-sections. The "lamellar" structure outside the pillar is the protective layer (95% Pt and 5% C) that was sputtered on the surface of the pillars before FIB milling to improve the flatness of the FIB cross-sections. The inner structure of the Ni-Mn pillars reveals a dense polycrystalline structure without voids, micropores, or other defects. The grain size of Ni-Mn is in the nanometer range, similar to the microstructure of the conventional electrodeposited Ni-Mn alloy [35]. From the EDX distribution of Ni and Mn over the whole cross-section (Fig.4(b)), it can be inferred that Ni and Mn are distributed uniformly over the entire feature. The inner structure of the Ni-Co pillar is also dense and polycrystalline (Fig.4(c)), which, in contrast to Ni-Mn, exhibits larger grains (up to micrometers across), which is also in accordance with that observed in conventional electrodeposited Ni-Co alloys [36]. Similar to the Ni-Mn alloy, Ni and Co were also found to be uniformly distributed over the entire feature (Fig.3(d)). High-resolution TEM images of the microstructure of the pillars as well as EDX analysis were carried out, and the results are shown in S2. This indicates the uniform distribution of alloy components in a single lattice in both the Ni-Mn and Ni-Co pillars. The combined evidence of the spatial metal distribution and the grain structure of Ni-Mn and Ni-Co, along with the similarity between the microprinted and conventionally electroplated-materials, suggests that the printed materials can be classified as solid solution alloys in which the Mn or Co atoms enter the Ni lattice [37,38].

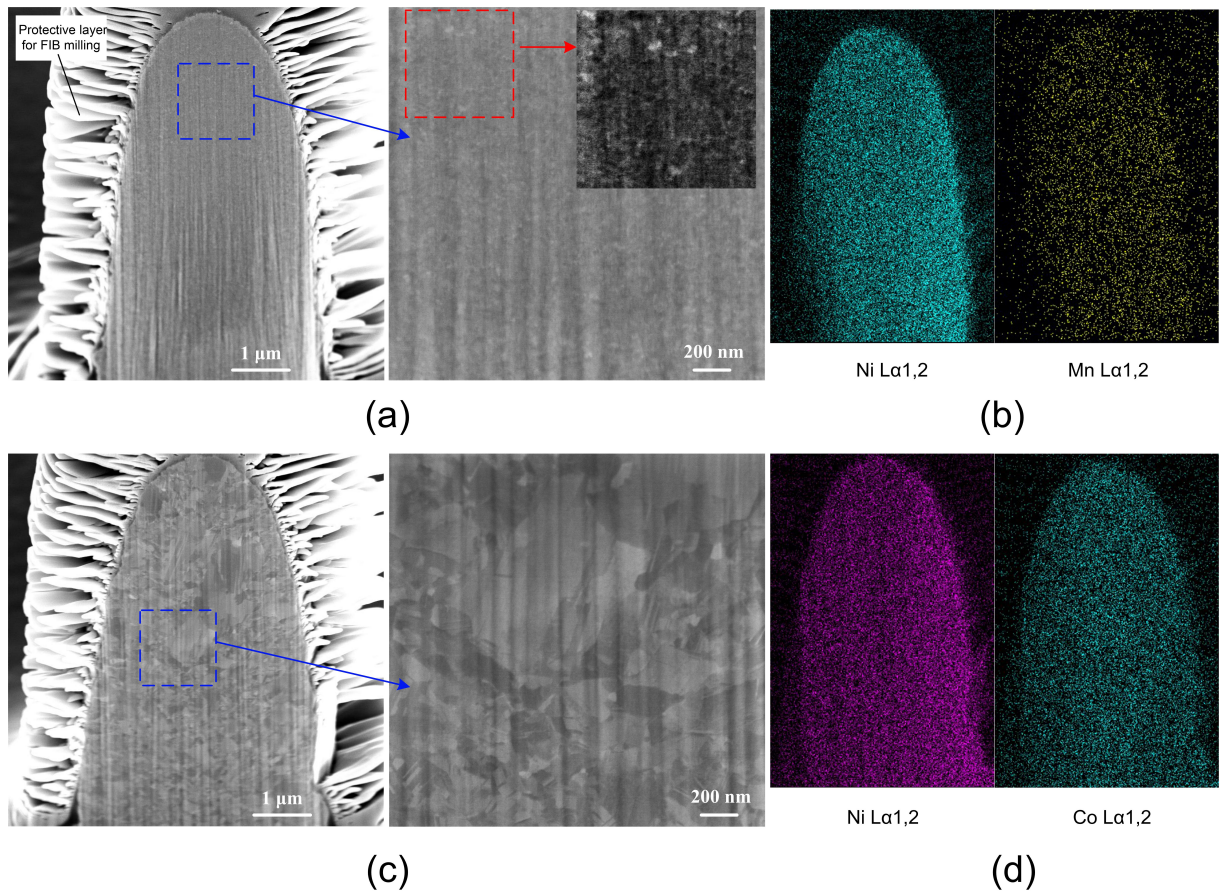

*Fig.4.* FIB-SEM cross sections of the pillars and distribution of alloy composition on the cross section: **(a)** cross section of a Ni-Mn pillar, **(b)** corresponding EDX distribution of Ni and Mn, **(c)** cross section of a Ni-Co pillar, **(d)** corresponding EDX distribution of Ni and Co.

## 4. Conclusions

In this study, Ni-Mn and Ni-Co pillars were prepared via local 3D electrodeposition with FluidFM. Printing was achieved by adapting previously developed metal-salt solution compositions for local electroplating. This was achieved by optimizing the printing potential and pressure for the localized ion delivery of metal-salt solutions containing $Ni^{2+}$ and one of the two other metal ions, $Co^{2+}$ or $Mn^{2+}$. The printing process was characterized by printing rates in the range of 50-60 nm/s. Characterization of the printed features by electron microscopy with the help of focused ion beam milling revealed the outer and internal structures of the deposits. The pillars revealed a smooth outer surface and a fully dense polycrystalline inner morphology. EDX analysis revealed a uniform distribution of metals with compositions in both the Ni-Mn and Ni-Co alloys, similar to macroscopically plated alloy thin films.

The electrodeposition of alloys is much different with pure metals, especially in FluidFM. This study demonstrated the potential of local microelectrochemical 3D printing via FluidFM to produce metal alloy features, which can be extended to the more complex microstructures required for many applications, such as alloy photonic crystals (e.g., Au-Ag), alloy interconnects (e.g., Cu-Ag, Pt-Cu), and alloy biosensors (e.g., Pt-Ni, Fe-Ni). Currently, printing is limited to alloys of fixed composition (given by the content of the metal-salt solution), but future extensions of the technology with multiple nozzles can, in principle, allow microfabrication with dynamically adjusted metal ratios in microprinted alloys.

## 5. Acknowledgements

The authors acknowledge financial support from the China Scholarship Council (No. 201706830034), Innosuisse (Swiss Innovation Agency) under project (18511.1 PFNM_NM), and Chinese Postdoctoral Science Foundation (2020M681580). D.M. acknowledges the funding from the European Research Council (ERC) under the European Union's Horizon 2020 research and innovation programme (Grant agreement No. 948238).

**Declaration of interests**

The authors declare that they have no known competing financial interests or personal relationships that could have appeared to influence the work reported in this paper.

**REFERENCES**

[1] Martin M L, Liew L A, Read D T, Christenson T R, DelRio F W and Geaney J 2020 Dominant factors for fracture at the micro-scale in electrodeposited nickel alloys Sens. Actuators A Phys. 314 112239.

[2] Guo Y H, Wang Y N and Xie L H 2011 Micro Electroformed Ni-P Alloy Parts by extended UV-LIGA Technology Adv. Mater. Res. 317–319 1635.

[3] Son S H, Park S C, Lee W and Lee H K 2013 Manufacture of μ-PIM gear mold by electroforming of Fe–Ni and Fe–Ni–W alloys Trans. Nonferrous Met. Soc. China 23 366.

[4] Yeh Y -M, Tu G C and Fang T -H 2004 Nanomechanical properties of nanocrystalline Ni-Fe mold insert J. Alloys Compd 372 224.

[5] Hsu H J, Huang J T, Lee K Y and Tsai T C 2012 Development of UV-LIGA contact probe, Microsystems, Packaging, Assembly and Circuits Technology Conference (IMPACT) 7th International (IEEE Publications).

[6] Ju X, Yang W, Gao S and Li Q 2019 Direct writing of microfluidic three-dimensional photonic crystal structures for terahertz technology applications ACS Appl. Mater. Interfaces 11 41611.

[7] Shen P, Wang G, Kang B, Guo W and Shen L 2018 High-efficiency and high-color-rendering-index semitransparent polymer solar cells induced by photonic crystals and surface plasmon resonance ACS Appl. Mater. Interfaces 10 6513.

[8] Su S et al. Reliability of micro-alloyed SnAgCu Based Solder Interconnections for various Harsh Applications 2019 69th Electronic Components and Technology Conference (ECTC) (IEEE Publications) (IEEE Publications).

[9] Watanabe A O, Wang Y, Ogura N, Raj P M, Smet V, Tentzeris M and Tummala R 2019 Low-loss additively deposited ultra-short copper-paste interconnections in 3D antenna-integrated packages for 5G and IoT applications 69th Electronic Components and Technology Conference (ECTC) (IEEE Publications) (IEEE Publications).

[10] Wang H, Zhang Y, Li H, Du B, Ma H, Wu D and Wei Q 2013 A silver–palladium alloy nanoparticle-based electrochemical biosensor for simultaneous detection of ractopamine, clenbuterol and salbutamol Biosens. Bioelectron. 49 14.

[11] Lv X, Ma H, Wu D, Yan T, Ji L, Liu Y, Pang X, Du B and Wei Q 2016 Novel gold nanocluster electrochemiluminescence immunosensors based on nanoporous $NiGd-Ni_2O_3-Gd_2O_3$ alloys Biosens. Bioelectron. 75 142.

[12] Wang H, Chen C, Yang F, Shao Y and Guo Z 2021 Direct ink writing of metal parts with curing by UV light irradiation Mater. Today Commun. 26. 102037

[13] Chang H, Guo R, Sun Z, Wang H, Hou Y, Wang Q, Rao W and Liu J 2018 Flexible conductive materials: Direct writing and repairable paper flexible electronics using nickel–liquid metal ink Adv. Mater. Interfaces 5. 187009717

[14] Merrow H, Beroz J D, Zhang K, Muecke U P and Hart A J 2021 Digital metal printing by electrohydrodynamic ejection and in-flight melting of microparticles Addit. Manuf. 37. 101703

[15] Park J U et al 2007 High-resolution electrohydrodynamic jet printing Nat. Mater. 6 782.

[16] Utke I, Michler J, Winkler R and Plank H 2020 Mechanical properties of 3D nanostructures obtained by focused electron/ion beam-induced deposition: A review Micromachines 11 397.

[17] Diercks D R, Gorman B P and Mulders J L 2017 Electron beam-induced deposition for atom probe tomography specimen capping layers Microsc. Microanal. 23 321.

[18] Hagen C W 2014 The future of focused electron beam-induced processing Appl. Phys. A 117 1599.

[19] Winhold M, Schwalb C H, Porrati F, Sachser R, Frangakis A S, Kämpken B, Terfort A, Auner N and Huth M 2011 Binary Pt-Si nanostructures prepared by focused electron-beam-induced deposition ACS Nano 5 9675.

[20] Lin J C, Yang J H, Chang T K and Jiang S B 2009 On the structure of micrometer copper features fabricated by intermittent micro-anode guided electroplating Electrochim. Acta 54 5703.

[21] Lin Y P, Zhang Y and Yu M F 2019 Parallel process 3D metal microprinting Adv. Mater. Technol. 4. 1800393

[22] Hirt L et al. 2016 3D microprinting: Template‐free 3D microprinting of metals using a force‐controlled nanopipette for layer‐by‐layer electrodeposition Adv. Mater. 28 2277.

[23] Reiser A, Lindén M, Rohner P, Marchand A, Galinski H, Sologubenko A S, Wheeler J M, Zenobi R, Poulikakos D and Spolenak R Multi-metal electrohydrodynamic redox 3D printing at the submicron scale 2019 Nat. Commun. 10 1853.

[24] Momotenko D, Page A, Adobes-Vidal M and Unwin P R 2016 Write-read 3D patterning with a Dual-Channel nanopipette ACS Nano 10 8871.

[25] Tseng Y T, Wu G X, Lin J C, Hwang Y R, Wei D H, Chang S Y and Pen K C 2021 Preparation of Co-Fe-Ni alloy micropillar by microanode-guided electroplating J. Alloys Compd 885. 160873

[26] Wang C, Hossain Bhuiyan M E, Moreno S and Minary-Jolandan M 2020 Direct-write printing copper–nickel (Cu/Ni) alloy with controlled composition from a single electrolyte using co-electrodeposition ACS Appl. Mater. Interfaces 12 18683.

[27] Ercolano G, Zambelli T, van Nisselroy C, Momotenko D, Vörös J, Merle T and Koelmans W 2020 Multiscale Additive Manufacturing of metal microstructures Adv. Eng. Mater. 22. 1900961

[28] Ercolano G, Nisselroy C, Merle T, Voros J, Momotenko D, Koelmans W and Zambelli T 2019 Additive manufacturing of sub-micron to sub-mm metal structures with hollow AFM cantilevers Micromachines 11 6.

[29] Yang Y F, Deng B and Wen Z H 2011 Preparation of Ni-Co alloy foils by electrodeposition Adv. Chem. Eng. Sci. 1 27.

[30] Yang N Y C, Headley T J, Kelly J and Hruby J M 2004 Metallurgy of high strength Ni–Mn microsystems fabricated by electrodeposition Scr. Mater. 51 761.

[31] Gamburg Y D and Zangari G 1963 Electrodeposition of alloys (Academic Press).

[32] Atanassov N and Mitreva V 1996 Electrodeposition and properties of nickel–manganese layers Surf. Coat. Technol. 78 144.

[33] Guo J, Guo X, Wang S, Zhang Z, Dong J, Peng L and Ding W 2016 Effects of glycine and current density on the mechanism of electrodeposition, composition and properties of Ni–Mn films prepared in ionic liquid Appl. Surf. Sci. 365 31.

[34] Bakhit B and Akbari A 2013 Nanocrystalline Ni–Co alloy coatings: Electrodeposition using horizontal electrodes and corrosion resistance J. Coat. Technol. Res. 10 285.

[35] Stephen A, Rossi F, Nasi L, Ferrari C, Ponpandian N, Ananth M V and Ravichandran V 2008 Induced ordering in electrodeposited nanocrystalline Ni–Mn alloys J. Appl. Phys. 103 C88.

[36] Omar I 2021 Electrodeposition of Ni-Co film: A review Int. J. Electrochem. Sci. 150962

[37] Yang Z B, Sun J, Lu S and Vitos L 2018 Assessing elastic property and solid-solution strengthening of binary Ni-Co, Ni-Cr, and ternary Ni-Co-Cr alloys from first-principles theory J. Mater. Res. 33 2763.

[38] Yokota M and Mitani R 1980 Interdiffusion in γ solid solution of the Ni-Mn binary alloy system Mater. Trans. 21 409.

# Supporting Information

# Electrochemical 3D Printing of Ni-Mn and Ni-Co alloy with FluidFM

Chunjian Shen, Zengwei Zhu, Di Zhu, Cathelijn van Nisselroy, Tomaso Zambelli,

Dmitry Momotenko

**S1 Morphology of Pillars during the printing process**

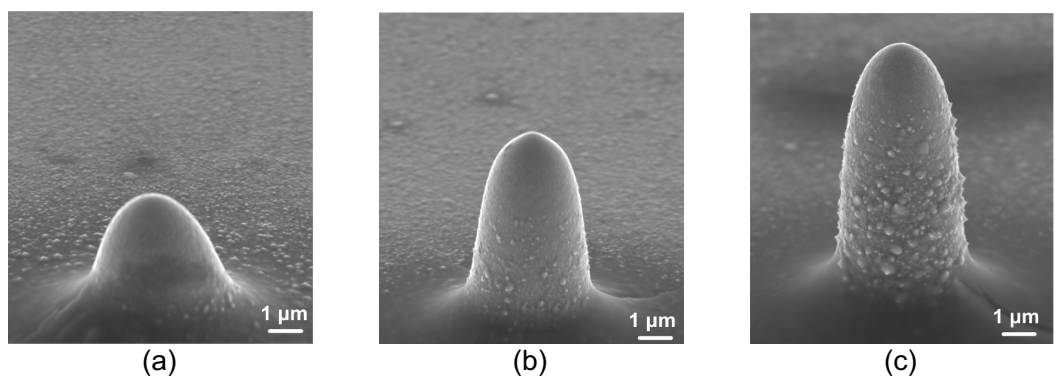

*Fig.S1.* SEM images of printed Ni-Mn pillars with 10 (a), 25 (b) and 40 (c) voxels

The Ni-Mn pillars with 10, 20 and 40 voxels are printed to show the changes of the pillars during the deposition process. The potential, the pressure, and the height of voxel of the pillar printing are -1.37 V, 8 mbar and 250 nm, respectively.

## S2 TEM images and EDX Analysis of the pillars

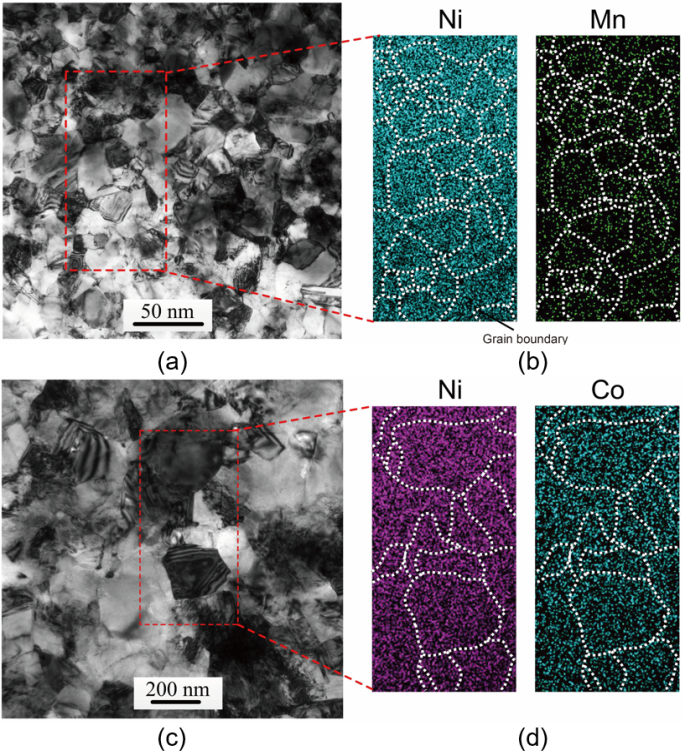

***Fig.S2.*** *TEM images of the pillars and distribution of alloy composition **(a)** TEM of a Ni-Mn pillar, **(b)** corresponding EDX distribution of Ni and Mn, **(c)** TEM of a Ni-Co pillar, **(d)** corresponding EDX distribution of Ni and Co. The painted white lines are used to show grain boundaries.*